\documentclass[modern]{aastex701}

\submitjournal{ApJ}

\usepackage{xfrac}

\shorttitle{Hostless Supernovae}
\shortauthors{Strolger et al.}
\begin{document}
\title{On the Origins of ``Hostless'' Supernovae: Testing the Faint-end Galaxy Luminosity Function and Supernova Progenitors with Events in Dwarf Galaxies}

\correspondingauthor{Louis-Gregory Strolger}
\author[0000-0002-7756-4440, gname=Strolger, sname=Lou]{Louis-Gregory Strolger}
\affiliation{Space Telescope Science Institute, 3700 San Martin Drive, Baltimore, MD 21218, USA}
\affiliation{Johns Hopkins University, Baltimore, MD 21218, USA}
\email{strolger@stsci.edu}

\author{Mia Sauda Bovill}
\affiliation{Department of Astronomy, University of Maryland, College Park, MD 20742 USA}
\email{msbovill@umd.edu}

\author[0000-0002-3099-1664]{Eric Perlman}
\affiliation{Department of Aerospace, Physics and Space Sciences, Florida Institute of Technology, 150 West University Boulevard, Melbourne, FL 32901, USA}
\email{eperlman@fit.edu}

\author{Craig Kolobow}
\affiliation{Department of Aerospace, Physics and Space Sciences, Florida Institute of Technology, 150 West University Boulevard, Melbourne, FL 32901, USA}
\email{ckolobow2014@my.fit.edu}

\author[0000-0003-2037-4619]{Conor Larison}
\affil{Department of Physics \& Astronomy, Rutgers, State University of New Jersey, 136 Frelinghuysen Road, Piscataway, NJ 08854, USA}
\email{cl1449@physics.rutgers.edu}

\author[0009-0003-8380-4003]{Zachary G. Lane}
\affiliation{School of Physical and Chemical Sciences$-$ Te Kura Mat\={u}, University of Canterbury, Private Bag 4800, Christchurch 8140, \\ Aotearoa, New Zealand}
\email{zachary.lane@pg.canterbury.ac.nz}

---------

\begin{abstract}

We present arguments on the likely origins of supernovae without associated host galaxies from open field, non-clustered, environments. We show why it is unlikely these ``hostless'' supernovae stem from escaped hyper-velocity stars (HVS) in any appreciable numbers, or the extreme outer halos of large galaxies, especially for core-collapse supernovae. It is highly likely that hostless events arise from dwarf host galaxies too faint to be detected in their parent surveys.  Several detections and numerous upper limits suggest a large number of field dwarfs, to M$_V>-14$, which themselves may be important to constraining the slope of the low-mass end of the UV luminosity function, understanding galaxy evolution, and putting $\Lambda$CDM into context. Moreover, the detailed study of these mass and metallicity-constrained host environments, and the variety of supernovae that occur within them, could provide more stringent constraints on the nature of progenitor systems. 
\end{abstract}


\section{Introduction} \label{sec:intro}

As extremely energetic stellar explosions, supernovae (SNe) generally reach brightnesses almost equivalent to the total integrated brightness of the galaxies they occur within. As such, it is rare to see SNe apparently unassociated with a host, but they do happen. There have been several dozen SNe discovered during the course of nearby, low-$z$ surveys, for which no host galaxy was identified to the detection limits of those surveys. It is likely that in most if not all cases, the underlying host galaxies for these `hostless' supernovae do indeed exist, and are an intrinsically faint population of dwarf galaxies with compact central regions, as faint as M$_{Rc}\ge-14$~\citep[see][Kolobow et al. in prep.]{Qin:2024il}. 

To be clear, there are SNe that occur in the intracluster environment of galaxy clusters~\citep{Gal-Yam:2003qf,Sand:2011fm, Graham:2015ik,Larison:2024dk}, wherein the intracluster light likely stems from a population of intracluster stars~\citep{Theuns:1997ez, Mihos:2016wa,Montes:2019pd,Montes:2022uf} stripped in numerous tidal interactions, and major and minor merger events. These are not the SNe we are concerned with in this manuscript. Rather, it is the hostless supernovae in non-clustered, ``open field'' environments, as such events could be used to provide information on intergalactic stellar populations (perhaps), confirm the faint end of the galaxy luminosity function, or test the progenitor mechanisms for some SNe types~\citep{Eldridge:2017dm}, as well as other investigations. For example, \cite{Lauer:2021xi} show, with data from the New Horizons Long-range Reconnaissance Imager, an excess optical ($\sim 6000$\AA) sky brightness of $\sim 10$ nW m$^{-2}$ sr$^{-2}$ in high galactic latitude fields, after accounting for zodiacal and galactic contributors. While it is, at present, unclear what diffuse or unresolved sources could be responsible for this cosmic optical background (COB), among the list of potential candidates are unresolved SNe apparently unassociated with host galaxies. In this paper, we attempt to determine what the contribution of these types of SNe might be, investigating some possible origins for these ``hostless'' SNe, as events from ejected intergalactic stars, as outer halo events, and as events in low-surface brightness or dwarf galaxies. 

\cite{Tyson:1987ue} postulated, based on the apparently high projected distance of SN~1983K~\citep{Niemela:1985dy}, far from the bright star formation regions of the host galaxy, NGC~4699, that ``extreme'' dwarf galaxies could host easily noticed SNe, yet themselves be too dim to be detected. Such SNe would appear to be similarly ``detached'' from any obvious host. While SN~1983K later proved to be accurately attributed to NGC~4699~\citep{Phillips:1990zf}, others have since taken up the challenge to look for these unassociated SNe. This is not an easy venture, as \cite{Hayward:2005rm} and others have shown, as such searches need not only depth but area to sample a sufficient volume to survey a number of such galaxies in a reasonable timeframe. 

In campaigns over two semesters in 2002 at \textit{Magellan} and the \textit{VLT}, L.~M.~Germany and L.-G.~Strolger undertook a project to recover the hosts of three such events from early nearby SN surveys, The Mount Stromlo Abell Cluster Supernova Search~\citep{Reiss:1998sw,Germany:2004tk}, and the Nearby Galaxies Supernova Search~\citep{Strolger:2003}. SN~1998bt~\citep{Reiss:1998sw} was shown to be a SN~1987A-like event in a host with M$_R=-12.6\pm0.9$. SN~1999aw~\citep{Strolger:2002pb} was a type Ia, although peculiar and SN~1999aa/91T-like, in a host with M$_B=-12.2\pm0.2$. In addition, 2000cd~\citep{Strolger:2000qf,Strolger:2003} was an apparently hostless narrow-line type II SN, with a long plateau phase similar to SN~1988Z. As such, no direct measure was made of the host brightness, but an upper limit of M$_R>-14$ was placed three years after the explosion.

Since then, several highly energetic supernovae have often been associated with low-luminosity, low-metallicity dwarf hosts galaxies, and even the least star-forming regions of those galaxies.  \cite{Lunnan:2014vo} have shown that type I superluminous supernovae~(SLSNe~I) often occur in low-metallicity hosts ($\approx 0.4 \,Z_{\odot}$) that are also low-luminosity ($\approx -17$ mag) and low-mass [$\log(M/M_{\odot})\approx 8$], while \cite{Hsu:2024if} further show SLSNe~I seem to be more offset from the light of their host galaxies than long GRBs \citep[e.g.,][]{Fruchter:2006kl}, than type Ic and Ic-bl~\citep{Modjaz:2020qp}, and than more normal CCSNe.\footnote{A caveat worth mentioning is that the comparisons to fractions of light in these studies were often not done in consistent rest-frame passbands for each SN type, or with significant accounting for extinction, and therefore the results may be muddled by different access to the attenuated star-formation rates.} This trend is well exemplified by SN~2016iet, an exceptional type I supernova and candidate pair-instability SN that is associated with a faint ($\approx-16$ mag) low-Z ($\approx0.1 \,Z_{\odot}$) and low-mass [$\log(M/M_{\odot})\approx 8.5$] host, albeit $\sim4$ $R_e$ from the center of the galaxy~\citep{Gomez:2019aa}. In fact, there is some evidence that it sits on a fainter knot which may itself be a dwarf satellite or ejected \ion{H}{2} region fainter than $\approx-15$ mag.
 
Conversely, there are SNe types that seem to be associated with the outermost halos of their hosts, both in location and along kinematic arguments. \cite{Foley:2015aa} show a trend for many Ca-rich SNe~Ib to be not only at large projected distances ($30-150$ kpc) from their host centroids, but also show evidence that the few without large offsets have large line-of-sight velocity shifts, relative to their hosts, implying they have been kicked to outbound trajectories by dynamical interactions. Along the same lines, some progenitors of other SNe have also been known to demonstrate unusually high velocities as well. For instance, the radial velocities of some luminous blue variables in the LMC are nearly 4$\times$ higher than red supergiants in the same region, suggesting luminous blue variables, which themselves become SNe~\cite[e.g.,][]{Smith:2011sr}, are likely kicked by the supernova of a companion star~\citep{Aghakhanloo:2022vz}. It remains to be seen if one such example would achieve or exceed the escape velocity of its host galaxy. There are also high-velocity white-dwarf stars, which themselves may be the survived companions of SNe~Ia. \cite{Shen:2018aa} have found at least three HVS that could have arisen from dynamically-driven double-degenerate double-detonation type Ia scenarios, at least one of which with an outbound velocity exceeding $\sim$1000 km/s.
 
As the breadth of large-scale surveys has vastly increased, so too has the number of hostless field SNe reported, from the Pan-STARRS PS1 and 3$\pi$ surveys~\citep{Chambers:2016jg,Flewelling:2020tu}, the Sloan Digital Sky Survey-II~\citep{Sako:2018wj}, the DESI Legacy Imaging Survey~\citep{Dey:2019vy},  the All-Sky Automated Survey for Supernovae~\citep[ASASSN, e.g,][]{Holoien:2019aa}, and the Zwicky Transient Facility Bright Transient Survey~\citep{Perley:2020vn}, to name a few, with hostless (galaxies fainter than M$_R>-14$) candidate samples numbering in the hundreds~\citep{Qin:2024il}. While these surveys have largely been relatively shallow (m$_g\la23$ to $5\sigma$), exchanging area for depth, that paradigm is about to change. With the launch and successful commissioning of \textit{Euclid}, the first light of \textit{Rubin}, and launch of \textit{Roman} in the next few years (hereafter, the Flagship Wide-area Surveys, or FWS), those survey volumes will greatly expand to expect several thousands of hostless SNe (and perhaps a few rarities from \textit{JWST} and \textit{HST}) to probe these questions on the nature of these hosts and their relations to SN progenitors by the end of the 2020s.  Indeed, recent followup of a sample of hostless SNe from the Pan-STARRS PS1 and ASASSN surveys has found a number of very faint dwarf hosts, including one with absolute magnitude M$_r=-12.7$ (Kolobow et al., in prep.).

In this manuscript we discuss the alternate hypothesis that these SNe originate from the escaped stars or outer-halo stars of large galaxies (Section~\ref{sec:2}), the expected rate of events from dwarf galaxies (Section~\ref{sec:dgxy}), how these hostless could test the faint-end galaxy luminosity function (Section~\ref{sec:4}), and provide constrained test-beds for SN progenitors (Section~\ref{sec:snprog}). Lastly, we discuss the potential contribution of unresolved hostless SNe to the \cite{Lauer:2021xi} COB (Section~\ref{sec:cob}).

\section{Hypervelocity stars as the progenitors of hostless supernovae}\label{sec:2}
Perhaps best articulated by \cite{Zinn:2011oz}, a question is whether a sufficient number of SN progenitor stars, as hypervelocity stars, would manage to get far enough away from their hosts before exploding to account for some significant fraction of hostless events? Or alternatively, are there enough low-mass dwarf galaxies in the field, not associated with clusters, with sufficient star-formation activity to give rise to a significant population of SNe? 

In the late 1990's and early 2000's, the term ``hostless supernova'' was most often connected with events that occur in clusters of galaxies. In testing the possibility they result from stars stripped in multiple interactions, and composing the intracluster light~\citep{Montes:2022uf}, \cite{Zinn:2011oz} argued whether or not there is sufficient time for a SN progenitor star (or system) to move far enough away from its host galaxy before it explodes as a supernova event. That time would depend on the distance the star or system would have to travel to ``escape'' from its host, its speed, and the time the system has before it would explode. By way of defining a criterion, many investigators adopted a projected distance of $\gtrsim20-30$ kpc $h^{-1}$ from the nearest visible edge of any galaxy in the detection imagery as the defining characteristic~\citep{Germany:2004tk,Sharon:2010,Dilday:2010a,Barbary:2012a,Graham:2015ik}, or essentially at a distance more than twice the visual radial extent of the nearest galaxy. This is convenient for these environments, given the density of potential hosts in rich clusters, and the abundance of potential progenitor stars abandoned in the intracluster medium. 

However, it is somewhat less useful as a criterion in field galaxies where the likelihood of any significant population of stars between galaxies is much smaller, due to much less frequent galaxy mergers~\citep{Jogee:2008mz,Husko:2022mq}. Moreover, the potential extent of stars bound to a galaxy extends broadly through the galaxy's dark matter halo, which can extend to distances $\sim5$ or more times the diameter of the visual extent~\citep{Deason:2020jn}. \cite{Gupta:2016ve} coined a more convenient term for this measure, the `directional light radius' (DLR), which is the radial extent of a neighboring galaxy's light\footnote{Derived from the Petrosian half-light radius, typically in the Sloan $r-$band.} in the direction of the event expressed in units of arcseconds~\citep[see also][]{Sullivan:2006a,Sako:2018wj}. The DLR distance, $d_{\rm DLR}$, is the number of DLR's away the SN is from the host nucleus. In this way, SNe at $d_{\rm DLR}>4$ or 5 from a potential host would be at distances in which $\ga95\%$ of the light is contained, and thusly $\le5\%$ of the stars are expected to lie, significantly minimizing the chance that the said SN could have originated from the given host galaxy. This represents an adjusted scale in projected distances dependent on the intensity profiles of the galaxies, where for large galaxies like the Milky Way (MW) with R$_e\simeq5-6$ kpc~\citep{Lian:2024du}, these $d_{\rm DLR}$ correspond to projected distances of $\sim20-30$ kpc and is much smaller for dwarf galaxies \citep[like Andromeda XIX with $\sim7-9$ kpc, ][]{McConnachie:2008id}. However, as can be seen in Figure 1 of  \cite{Qin:2024il}, nearly all hostless SN discoveries have significantly larger projected distances from any nearby MW-sized galaxies, closer to $\sim100s$ of kpc than dozens of kpc.

\subsection{Could a potential SN progenitor star get far enough away?}
In this section we assess if a given SN progenitor star system would have sufficient time to achieve distances of $\sim100s$ kpc before they would explode, assuming the stars originate in the inner half-light radius of their hosts. Most stars in the MW, for instance, move along at well below the galaxy's escape velocity of $v_{\rm esc}\simeq550$ km/s~\citep{Kafle:2014qt}. Hypervelocity stars (HVS), while rare, have been known to greatly exceed this speed limit, with $v\gtrsim1000$ km/s~\citep[see][on recent discoveries of nearby high-velocity metal-poor L subdwarfs]{Burgasser:2024aa,Zhang:2018aa}.  Potentially the results of dynamical ``kicks'' from intermediate-mass black holes within globular clusters~\citep[e.g.,][]{Cabrera:2023aa}, or possibly the ejected companions of SNe themselves~\cite[e.g.,][]{Shen:2018aa}, these stars would be capable of traversing extraordinary distances, at speeds of $\sim 1$~kpc/Myr. If on out-bound orbits, these HVS could easily reach distances of 100s of kpc in an equivalent number of Myr. Using this speed as a guide, assuming one's definition of unassociated distance is $\gtrsim100$ kpc, and setting $t_{\rm esc}\gtrsim100$ Myr, is that significantly longer than the timescale to producing SNe, or is $t_{\rm SN} \gg t_{\rm esc}$?

For massive-star core-collapse SNe (or CCSNe), $\gtrsim90$\% of the time to go from star-formation to explosion ($t_{\rm SN}$) is consumed in the time on the main-sequence ($t_{\rm MS}\approx t_{\rm SN}$). That main sequence lifetime is well approximated by the nuclear burning timescale and the mass-luminosity relation for upper main sequence stars, or $t_{\rm MS}\approx 10\, (M/M_{\odot})^{-2.5}$ Gyr.  That corresponds to lifetimes of $\lesssim30$ Myr for stars $\gtrsim10\,M_{\odot}$, about 3 times less than the timescale needed for an HVS to reach 100s of kpc from the host. It would appear rare for such progenitors to reach such distances.

On the other hand, white dwarf supernovae, or type Ia SNe (or SNe~Ia) have an additional delays in $t_{\rm SN}$ than just their progenitor main-sequence lifetimes. The current convention is that the majority of these result from the mergers, or near mergers, of white dwarf (WD) binaries, and are governed by the time necessary to radiate angular momentum from the merging system~\citep{Maoz:2011,Strolger:2020aa}, affected by the pair's initial separation. These double-degenerate (DD), WD-WD mergers thusly display a wide distribution of delay times, equivalent to $t_{\rm SN}$ used here, that could be described equally well by a power-law distribution, with $\Phi(t)\propto t^{-\beta}$ where $\beta\approx1$ and is truncated below $\sim30-50$ Myr~\citep{Rodney:2014fj}, or by an exponentially declining distribution, $\Phi(t)\propto\exp(-t)$~\citep{Strolger:2020aa}. In either, the average delay-time is $t_{\rm SN}\sim700-800$ Myr from formation to explosion, where many ($\gtrsim 40\%$) have $t_{\rm SN} \gtrsim 1$ Gyr. This would be more than sufficient time for even bound stars on eccentric orbits at the average velocity to reach distances of a few 100s of kpc from the galactic center. 

It should also be pointed out that a virialized stellar population in equilibrium should approach a Maxwell-Boltzmann (M-B) distribution of velocities, which is expected for a thermalized gas.  While galaxies are not closed systems, it is instructive to estimate how rare such systems are using this conceptualization.  The velocity dispersion of stars should be the critical factor here.  So, for example, the velocity dispersion of the disk in the Solar neighborhood according to Gaia data is $\sim 40$ km/s~\citep{Gaia-Collaboration:2018ab}.  If this is taken as the peak in the M-B distribution, then for a galaxy of $10^{12}$ stars we would not expect even one such star, as the probability of a star with $v > 550$ km/s would be $\sim 10^{-40}$.  This indicates that the progenitor, high velocity stars would need to be associated with a different stellar population that is higher velocity dispersion and likely older and/or higher in the disk~\citep[see, e.g., ][]{Sharma:2021aa}.

It would seem that while it is not likely HVS to survive the trip long enough to be the progenitors of hostless CCSNe, it would be clearly possible for HVS-SNe~Ia to exist, at least on the basis of approximate timescales.

\subsection{The rarity of HVS}

There are 1.8 billion stars in the \textit{Gaia} DR3 release~\citep{Gaia-Collaboration:2023sj}, $\sim600$ are now known high-velocity stars in the MW, $\sim50$ of which are of the hyper-velocity type, and perhaps only $\sim5$ with a 50\% chance or greater of escaping the Galaxy~\citep{Li:2020vu}.

If these numbers, approximately 1 escaping HVS in timescales of 100 Myr per 400 million stars, are representative of the fractions of HVS in other galaxies, it would appear HVS progenitors are too rare to attribute to hostless SNe, as the rate of discovery of hostless SNe is on the rise (see Section~\ref{sec:dgxy}), now on order of a couple of dozen per decade. While it is hard to normalize this event rate in the context of escaped stars from galaxies, it is presumably too frequent to be fully attributable to escaping high velocity stars.

With the exception of the two recent metal-poor L subdwarf discoveries, at present most HVS discoveries have been of massive, early type stars~\citep{Li:2020vu}, B-type or earlier. While this could be a brightness selection bias, it supports the nature of their possible origin, resulting from binary-system break up by dynamical interactions.

As earlier stated, \textit{Gaia} could have a selection bias against the faint progenitors of SNe~Ia, but there are approximately 360,000 WD in DR3~\citep{Gentile-Fusillo:2021ke}. Approximately 150,000 of these are fairly nearby ($d<500$ pc), and very few of have $v>100$ km/s~\citep{Mikkola:2022zz}. A deeper inspection of the brighter, younger, and more massive ($0.5M_\odot<M\lesssim1 M_\odot$) DA WDs in $d<100$~pc show only a fraction ($\sim 5\%$) with $v\gtrsim100$ km/s~\citep{Kilic:2020km}. In the more local neighborhood of $d<50$~pc, there are $\sim30$ known WDs, approximately 20 with proper-motion companions, of which most are main-sequence companions, but there is only one that is a double WD system, each with approximately the same age~\citep{Golovin:2024hk}. So it would seem \textit{Gaia} is sensitive to binary high or hyper-velocity WDs, but has thus far only identified on order 1 plausible system.

Unlike the D$^6$ SN~Ia progenitor scenario~\citep{Shen:2018aa} where the surviving high-velocity WD is the potential outcome of a SN, the progenitor pair would have to travel together at high-velocity to successfully complete mass transfer and explode at distance. Most dynamical studies of such three-body (or greater) encounters in dense globular clusters, or the galactic center, result in cataclysmic events and single stellar dynamical ejections~\citep{Cabrera:2023aa,Leonard:1991fu}, and not ejected pairs. Such dense environments can, however, increase the WD+WD merger rate, but not in appreciable numbers (by orders of magnitude) to account for much of the observed SN~Ia rate~\citep{Hamers:2013zp,Toonen:2018sh}.

It would seem HVS are too rare in occurrence and generally lack the necessary speed to attribute to hostless CCSNe. It is also not expected that SN Ia progenitor systems would survive the dynamical kick to give them high- or hyper-velocities, although again, it is not necessary that they have such speeds to reach the outer halo.

\subsection{What about outer halo stars?} 
SN~Ia progenitor systems on out-bound trajectories could sufficiently achieve $d\sim100$ kpc in a few Gyrs, which is sufficient timescale for $\gtrsim\sfrac{1}{3}$ of events by delay-time distributions without needing to be HVS to do so. Such scenarios are also consistent with the high-mass double-white dwarf transverse velocity dispersions, which are also the older WDs, allowing for a sufficient merger/interaction timescale~\citep{Cheng:2020jy}. But it is also very possible for WD+WD pairs in the outer halo, born in situ, to result in events mistaken as hostless.

The MW's outer-stellar halo extends beyond $d\gtrsim100$ kpc~\citep{Kafle:2014qt,Sharma:2011aa}. \cite{Ruiter:2009ao} assessed the number of close-proximity halo WD binaries with orbital periods $\lesssim 5$ hrs, as potential background sources to the \textit{Laser Interferometer Space Antenna} (\textit{LISA}) gravitational radiation experiment. Assuming a MW halo stellar mass of $\sim10^8-10^9$ $M_{\odot}^{-1}$~\citep{Bell:2008zk, Deason:2011ws,Deason:2019xw}, with $\eta_B\approx0.3$, simulations by \cite{Ruiter:2009ao} determined the number of such binaries is $\sim 10^6-10^7$. Assuming these WD+WD systems explode along the distribution of delay times ($\Phi$) with average times, as $\bar{t}=\int t \Phi(t)\,dt/\int \Phi(t)\,dt$, of $\bar{t}=2-5$ Gyr, consistent with~\cite{Iben:1998wj} on the WD+WD merger times from orbital periods, it implies a SN~Ia rate in the MW halo of $2-4\times10^{-4}$ yr$^{-1}$. Similar conclusions can be reached from an analysis of massive WDs from \textit{Gaia} as products of WD mergers, in which \cite{Cheng:2020jy} suggest a MW binary WD merger rate of $\sim10^{-13}$ yr$^{-1}$ $M_{\odot}^{-1}$. It could also be reached by approximating the SN~Ia rate as the product of two environmental components~\citep[see][]{Andersen:2018dp}, one which scales with star-formation rate, and the other with stellar mass, and only considering the mass term, with $B=4.7\pm0.6\times10^{-14}$ SNe~Ia yr$^{-1}$ $M_{\odot}^{-1}$. This would be consistent with the environments of 91bg-like SNe~Ia and Ca-rich SNe~Ia~\citep{Qin:2024aa}. When applying either to lower estimates for MW stellar halo, both result in SN~Ia rate in the MW halo of $\sim10^{-5}$ yr$^{-1}$.

Adopting this latter rate, and assuming these numbers are representative for L$_{\star}$ galaxies and brighter, and further assuming the local number density follows a \cite{Schechter:1976uq} analytic function with the number of such galaxies to $d<500$ Mpc about 2 million, the local SN~Ia rate from halos of bright galaxies would be $\sim 50$ yr$^{-1}$, only a factor of a few larger than what is currently observed in the local universe (see Section~\ref{sec:snprog} and Figure~\ref{fig:2}). However, if one adopts the former SN~Ia rate from \textit{LISA} WDs in the MW halo of $2-4\times10^{-4}$ yr$^{-1}$, it yields $\sim500$ SNe~Ia yr$^{-1}$, which is more than an order of magnitude larger than what is currently observed. The excess is likely due to the concentration of halo light, and presumably stars, toward the inner-halos of these galaxies, such that in M31 (for example) approximately 90\% of the light is contained interior to 20 kpc~\citep{Courteau:2011wk}. SNe~Ia originating from the inner halos would more likely be attributed to their hosts, as with events from the bulge or disk components of galaxies, and not likely be mistaken as hostless events. If these outer-halo environments are indeed responsible for the hostless SN~Ia population, then there may be additional factors on survey completeness that need to be more accurately accounted for in the estimate of the local SN rates from the recent low-$z$ surveys, or will be more precisely understood in the FWS era.

The massive stellar progenitors of CCSNe are estimated to be only a tiny fraction of halo stars, and likely migrated there on high (or hyper velocity) eccentric orbits, Therefore, it is not expected that CCSNe in stellar halos occur at a rate that could be mistaken for the hostless CCSN rate. Another important fact is that the SN rate per normal galaxy, at least in the local universe, has been firmly established to be about 2-10 per century \citep{vandenbergh1983388}. For most SNe, event rates are directly tied to regions of star formation within a galaxy more so than the regions with the highest number of stars, although there is undoubtedly a secondary correlation~\citep{Scannapieco:2005,Andersen:2018dp}. For those reasons, it is typically expected these hostless SNe originate from unseen faint or low-surface brightness stellar populations, with some active star formation, rather than the few stars remaining or ejected from their age of formation.

\section{SNe in Dwarf Galaxies}\label{sec:dgxy}
With the advent of ``all-sky'' surveys such as Pan-STARRS and ZTF, there have been over a hundred reported discoveries in just the last decade~\citep{Qin:2024il,Pessi:2024zw}. While their faint hosts are typically beyond the detection threshold of their discovery surveys, concerted followup has often resulted in direct detection of a faint dwarf galaxy, within only a few projected kpcs of the SN locations, in the absolute magnitude range $M_V\gtrsim-14$~\citep{Strolger:2002pb,Prieto:2008,Zinn:2012gf}. 

The volumetric relationship of the rate of CCSNe to the cosmic star-formation rate is often expressed as,
\begin{equation}
	R_{CC}\cong k_{CC}\,\dot{\rho}_{\star},
	\label{eqn:1}
\end{equation}
where the CCSN rate density, $R_{CC}$ in yr$^{-1}$ Mpc$^{-3}$, is directly related to the star formation rate density, $\dot{\rho}_{\star}$ in $M_{\odot}$ yr$^{-1}$ Mpc$^{-3}$, scaled by the fraction of the initial mass function (IMF) that give rise to core-collapse supernovae, $k_{CC}$ in $M_{\odot}^{-1}$. The scaling can be calculated from a Salpeter-like IMF, or derived from observations. There is a similar relationship for SNe~Ia,
\begin{equation}
	R_{Ia}= \varepsilon\,k_{Ia}[\Phi \ast \dot{\rho}_{\star}](t),
	\label{eqn:2}
\end{equation}
which includes two additional terms, an exponential or power-law delay-time distribution, $\Phi(t)$, of DD-WD systems which is convolved with the cosmic star-formation rate, and a mechanism efficiency, $\varepsilon$, accounting for the fact that far from all stars which become WDs explode as SNe~Ia. 

A way to determine the contribution to the volumetric star-formation rate from dwarf galaxies would be to sum up the product of the number density of such galaxies, and the typical star-formation rates (SFR) per galaxy, by
\begin{equation}
	\dot{\rho}_{\star}=\int_{\rm M_{V,\rm min}}^{\rm M_{V,\rm max}} n(\rm M_V)\,\,{\rm SFR}(\rm M_V)\,\, d{\rm M}_V,
	\label{eqn:3}
\end{equation}
where, M$_{V,\rm min}\ge-5$ and M$_{V,\rm max}\le-14$. The low-end slope of the galaxy luminosity function,  $n({\rm M}_V)$, is however not confirmed to these absolute magnitudes, but extrapolation suggests there should still be a healthy population of faint dwarfs following the $\alpha=-1.1\pm0.2$ tail of the power-law distribution~\citep{Schechter:1976uq}.  

\begin{figure*}[t] 
   \centering
   \includegraphics[width=\textwidth]{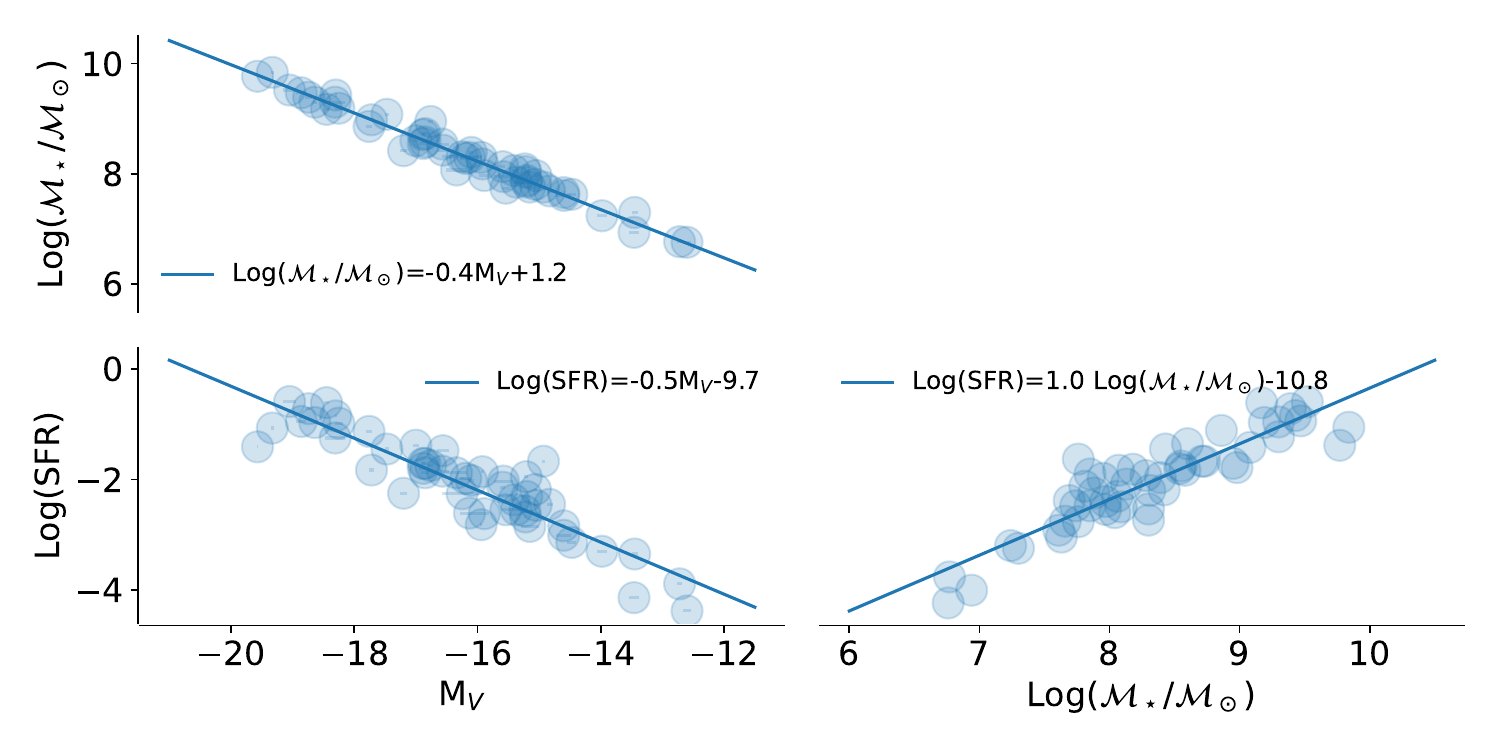}
   \caption{\footnotesize Star-forming main sequence of dwarf and low-surface brightness galaxies showing the relationships between stellar mass (in $\mathcal{M}_{\odot}$), central $V$-band absolute magnitude, and star-formation rates ($\mathcal{M}_{\odot}\, yr^{-1}$). Reproduced from \cite{McGaugh:2017tm}.}
   \label{fig:1}
\end{figure*}

The relation between the absolute magnitudes, stellar masses, and star-formation rates of dwarf galaxies has been well established~\citep{Schombert:2011vt, McGaugh:2017tm}, and is reproduced here in Figure~\ref{fig:1}, showing a nearly linear relationship between $\log$(SFR) and M$_V$ (\textit{lower left panel}).
Integrating the product of this SFR$-$M$_V$ relationship with the Schechter function yields an expectation of the star-formation rate density from dwarf galaxies, from Equation~\ref{eqn:3}, and expected SN rates from dwarf galaxies, from Equations~\ref{eqn:1} and \ref{eqn:2}, both shown in Figure~\ref{fig:2}.

\section{Testing the faint-end of the galaxy luminosity function}\label{sec:4}
One of the more intriguing applications of hunting the dwarf-galaxy hosts of hostless SNe is for completing the general knowledge of the luminosity function of galaxies at low-$z$. As discussed in \cite{Conroy:2015xy}, the low luminosity and low-surface brightness of field dwarf galaxies, as well as ultra-diffuse field galaxies~\citep{Bovill:2009qc,Bovill:2011dz,Bullock:2010jx}, with masses well below $<10^6\,M_{\odot}$, are expected to be numerous, yet only a handful of such field dwarfs have been detected to date~\citep{Hunter:2006qy, Chiboucas:2009cq, Karachentsev:2013cl}. A large reason is there has not been a survey with sufficiently large etendue, accounting for both their intrinsic faintness the sparse spacial density, to reasonably probe the population of field dwarf galaxies. The \textit{Rubin} Observatory's LSST should be capable of probing the density of such galaxies, to distances within a few 100s of Mpc, perhaps, with the deep 10-year co-added images. But until then, SNe provide ``signposts'' for locating dwarfs, independent of the wider environments of these galaxies.

Understanding how stellar masses, star-formations rates, and halo masses are related and evolve with time are important to understanding the $\Lambda$CDM paradigm. The evolution in the galaxy stellar mass function alone has few constraints on $\alpha$, even at low $z$, despite showing some evidence of evolution, to $\alpha\simeq-2$ at $z\sim8$~\citep{Navarro-Carrera:2024ch}. Advances will come as more field dwarf galaxies are identified, at $\mu\gtrsim27$ mag arcsec$^{-2}$, to significantly larger distances ($d_C\simeq1$ Gpc), where evolution can be accurately quantified~\citep{Conroy:2015xy}. Further understanding will come with reaching the stellar resolution of these galaxies, enabling resolved-star color-magnitude diagram fitting techniques, which through targeted followup with the MICADO instrument for the \textit{ELT}, and similar instrumentation on the \textit{GMT} or \textit{MMT}, should be achievable~\citep{Michaowski:2021oe}.

As shown in \cite{Conroy:2015xy}, the SNe that occur within these environments are expected to be numerous, on the order of hundreds per year to $d_C<1$ Gpc, providing ample ``signposts'' of where these dwarf galaxies are to enable targeted followup. By revealing the locations of dwarf galaxies for more targeted deep followup, the potential to contribute to the study of nearby low-mass galaxies is strong, and has implications on the nature of dark matter, cosmic reionization, and galaxy formation via ``near-field cosmology". Increasing the number of known faint (M$\gtrsim -12$) and ultra-faint  (M$\gtrsim -8$) dwarf galaxies would necessitate extending the matter (dark matter) power spectrum to well below $10^9\,M_{\odot}$ \citep{Bullock:2017ai, Buckley:2018oq, Jethwa:2018xh}. Further, resolving local dwarf galaxies, and creating an accurate census of the star-formation histories, may be the only way to link the faintest galaxies to reionization, and constrain the faint end of the UV luminosity function in the early universe \citep{Robertson:2015nq, Boylan-Kolchin:2015wp, Weisz:2017sr}.

\begin{figure*}[t] 
   \centering
   \includegraphics[width=\textwidth]{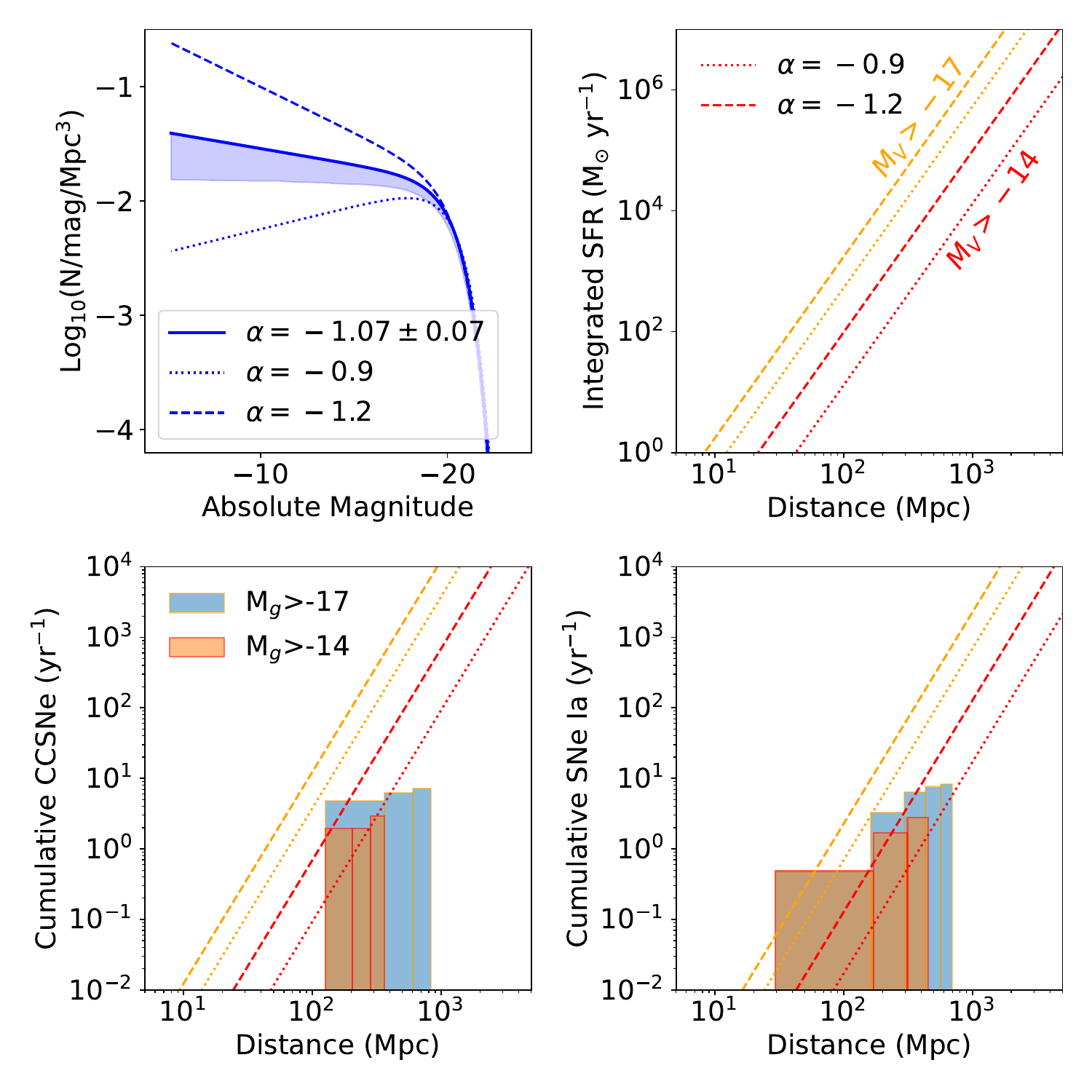}
   \caption{\footnotesize Predicted supernova rates (lower panels): from integrated galaxy luminosity functions, shown in upper left. Upper right shows the integrated star-formation rates by distance for the two extreme $\alpha$ (indicated), from dwarf galaxies fainter than -17 (in yellow) and -14 (in red) mag, respectively. Lower panels show the cumulative SN rates for CCSNe (lower left) and SNe Ia (lower right) with co-moving distance, with bars indicating cumulative rates to date, inferred from \cite{Qin:2024il}.}
   \label{fig:2}
\end{figure*}
The rates of SNe themselves may be useful for constraining the faint-end slope of the Schechter function. Figure~\ref{fig:2} shows the predicted cumulative rate of SNe (CCSNe in lower left; SNe Ia in lower right) from galaxies fainter than M$_V>-17$ (in yellow), and even fainter still at M$_V>-14$ (in red), in events per year over the entire sky. These are shown for two different assumed slopes of the dwarf galaxy luminosity function, as indicated by the dashed and dotted lines in the legend of the upper-left panel of Figure~\ref{fig:2}. Also shown are the cumulative rates to the same two magnitude limits, inferred from the \cite{Qin:2024il} sample of low-$z$ events over a 5.5-year period. For simplicity, we assume the bulk of these events stem from routine monitoring with large-area surveys, covering  $\sim1/2$ of the sky. We also assume that those surveys are complete, i.e., no events bright enough to be detected were missed either due to cadence gaps, or extraordinary line-of-sight extinction. 

It is interesting that with those basic assumptions, the observed cumulative rates are indeed approaching the expected yields for dwarf hosts, albeit perhaps a magnitude or two lower in total number for the respective absolute magnitude limits. Prior to \cite{Qin:2024il}, there simply was not a uniform sample with which to test the cumulative rates in these environments, and the remaining shortfall is likely in the details of survey cadences, coverage on sky, and exact sensitivities. It is very likely that with FWS, and perhaps a large-effort survey with \textit{JWST}, more SNe in these environments will be found, and more detailed analyses will be performed, closing the gap in expected vs. observed rates and providing valuable constraints on the value of $\alpha$. 

\section{Low-mass, low-metallicity host galaxies as constrained testbeds for SN progenitors}\label{sec:snprog}
Chemical evolution, particularly in Mn abundances, has been shown to be different in the MW satellites than for our own Galaxy, possibly indicating a separate dominant channel for SN~Ia production in these dwarf galaxy environments~\citep{Kobayashi:2015xo,Sanders:2021ik}. If the pathways for the Mn-deficient channels (e.g., sub-Chandrasekhar mass CO WDs) require systematically shorter delay-times than their Mn-rich counterparts~\citep{Kobayashi:2015xo}, methods to recover delay time distributions in different environments may be the key to distinguishing dominant progenitor mechanisms.

Dwarf galaxies in general may also have simpler star-formation histories than more normal galaxies~\citep{Weisz:2014jc}, providing a unique environment to probe progenitor relationships with environmental star-formation rates, masses, and metallicities. The results of star-formation driven outflow studies \citep[e.g.,][]{Romano:2023wu} show that enriched interstellar gas from the shallower gravitational potential wells of dwarf galaxies is driven out before it can be turned into more metal-rich stars. In turn, the connection between delay-time distributions and star-formation histories, as outlined in Section~\ref{sec:dgxy} volumetrically, is also applicable to the reconstructed star-formation histories of the galaxies themselves~\citep{Joshi:2024cr,Strolger:2020aa}, whether done through resolved-star CMD fitting~\citep{Hidalgo:2017oh}, or stellar population inference~\citep{Johnson:2021zh}. 

\section{The potential integrated contribution of SNe in Dwarf Galaxies to the Cosmic Optical Background}\label{sec:cob}
As earlier indicated, unresolved SNe have been postulated as potential contributors to the  \cite{Lauer:2021xi} COB. 
With the SN rates above, we can estimate the contribution of such events, following,
\begin{equation}
B(z) = \frac{1}{4\pi}\int_0^z W\cdot  \bar{F}_{\rm SN}(z')\cdot R_{\rm SN}(z')\,dV(z'),
\end{equation}
where, for a given SN type, $W$ is the window, or the approximate fraction of a year that SN would be detectable, approximately 30 rest-frame days for SNe~Ia and 100 days for CCSNe. Over that window the SN type is at an average luminosity, $\bar{L}$, derived from \cite{Richardson:2014fk} absolute magnitudes, and \cite{Kessler:2009a} lightcurve templates, which corresponds to a received flux of $\bar{F}_{\rm SN}(z)=\bar{L}\, [4\pi D^2_L(z)]^{-1}$ for SNe at redshift $z$. The product of these, with the appropriate SN rate and integrated with volume, yields the time-averaged COB light from that SN type.
\begin{figure*}[t] 
   \centering
   \includegraphics[width=\textwidth]{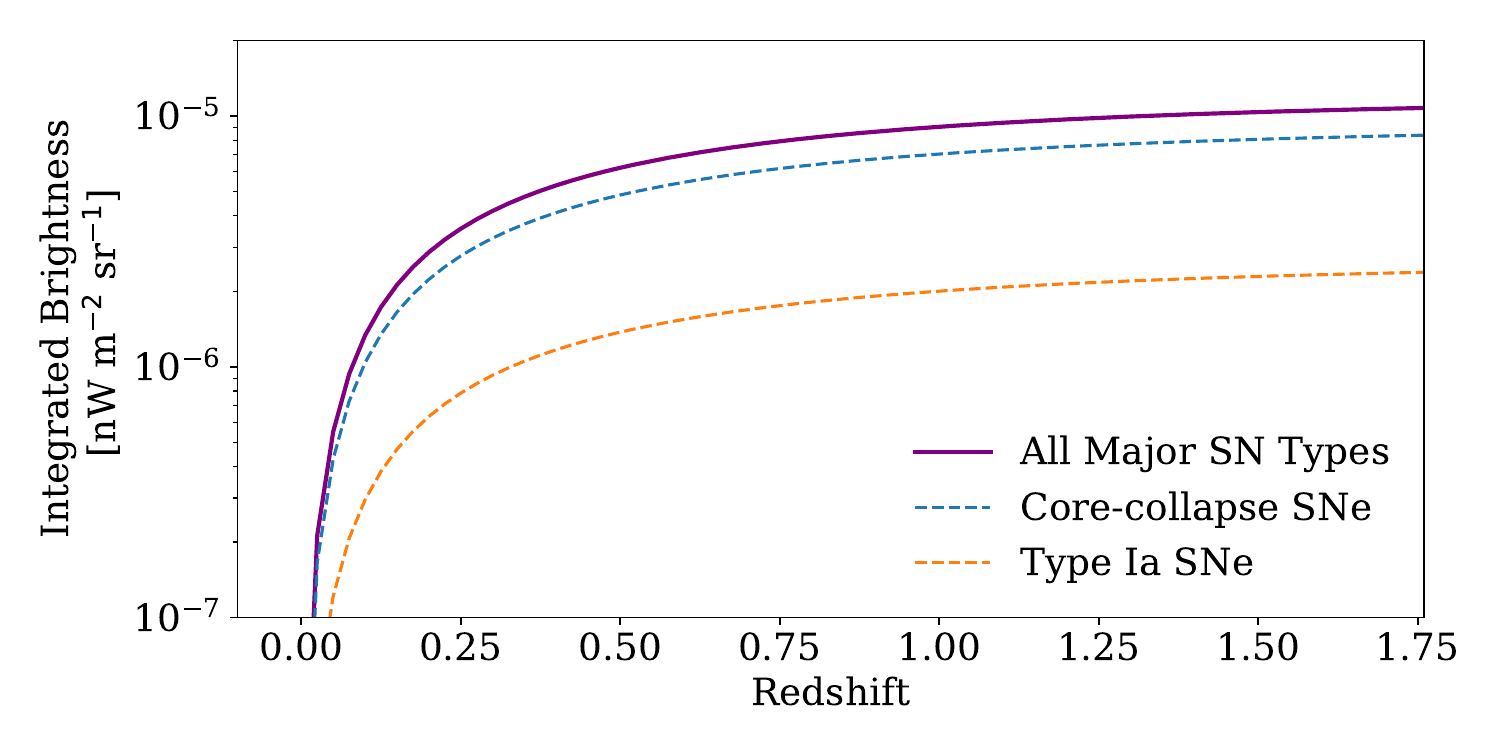}
   \caption{Approximated integrated contributions of unresolved SNe, from CCSNe and SNe~Ia, to the COB.}
   \label{fig:3}
\end{figure*}

Figure~\ref{fig:3} shows an approximation of the contribution of SNe to the COB. Here, to provide only an approximate calculation, we neglect many secondary redshift-dependent effects such as time-dilation which increases window function with redshift, $K$-corrections as rest-frame optical light shifts out of the observed optical passband, further reducing their apparent flux. For simplicity, we also do not account for the evolution in cosmic star-formation rate density~\citep{Madau:2014fk}, which for normal galaxies also increases the SN rate density with redshift by about an order of magnitude, likely peaking at $1<z<2$~\citep{Strolger:2020aa}, but it is also important to note that dwarf and low-mass galaxies may not show a  significant increase in star-formation rate density~\citep{Davies:2009ja,Cedres:2021xi}. A deeper quantitive or numerical analysis could be done to arrive at a more precise assessment of the contribution, if desired. But as can been seen in Figure~\ref{fig:3}, the SN contribution is already estimated to be $5-6$ orders of magnitude fainter than the measured values from \cite{Lauer:2021xi}.

\section{Summary}
We have presented arguments as to why it is unlikely hostless SNe stem from HVS in any appreciable numbers, for CCSNe in explosion time scales and for SN~Ia progenitor systems in surviving dynamical kicks. Further, we show that SN~Ia in the outer halos of large galaxies could account for the existence of hostless SNe~Ia. However, attributing hostless SNe to dwarf galaxies shows better agreement with observed SN rates for both SN types. This is fortunate, as these environments can be used as metallicity and mass-constrained testbeds for SN progenitor scenarios. 

Getting just the sheer number density of these dwarf galaxies will be important to understanding the formation history of low-mass galaxies in the universe, placing useful constraints on $\alpha$, the UV luminosity function, and $\Lambda$CDM, which may not be feasible until the first complete data releases of \textit{Euclid}, \textit{Roman}, \textit{Rubin}/LSST. Until then, the SNe produced within them tell us where these galaxies are, allowing for more targeted, deep observations. 

If the COB is a real phenomenon, not attributed to \textit{Kepler} instrumentation, it is unlikely hostless SNe extending out to the very early universe contribute much in their integrated light. 

\begin{acknowledgments}
We thank our anonymous referee for valuable comments and insights which have improved this manuscript. We thank Dr.~Lisa M.~Germany for her early pioneering work in investigating SNe in dwarf galaxies. LGS also thanks Dr.~Sebastian Gomez for his valuable edits and contributions. CL acknowledges support from DOE award DE-SC0010008 to Rutgers University.
\end{acknowledgments}
{\large\it Author contributions:} We use the CRT standard (\url{https://authorservices.wiley.com/author-resources/Journal-Authors/open-access/credit.html}) for reporting author contributions. Conceptualization: L.G.S. Data curation: L.G.S. Formal analysis: L.G.S. Investigation: L.G.S., M.S.B., E.P., C.K. Methodology: L.G.S. Software: L.G.S. Supervision: L.G.S. Validation: L.G.S. Writing$-$original draft: L.G.S. Writing$-$review \& editing: L.G.S., M.S.B., E.P., C.K., C.L., Z.G.L.

\bibliography{hostless}{}
\bibliographystyle{aasjournalv7}
\end{document}